\documentclass[%
 reprint,
 amsmath,amssymb,
 aps,
]{revtex4-2}

\usepackage[dvipsnames]{xcolor}
\usepackage{graphicx}
\usepackage{dcolumn}
\usepackage{bm}
\usepackage{tikz}
\usepackage{pgfplots}
\usepackage{subfigure}
\usepackage{multirow}
\usepackage{diagbox}
\definecolor{babyblue}{RGB}{131,195,221}
\definecolor{lightblue}{RGB}{48,155,200}
\definecolor{justblue}{RGB}{1,102,169}
\definecolor{deepblue}{RGB}{2,52,107}

\definecolor{babyred}{RGB}{254,134,110}
\definecolor{lightred}{RGB}{252,67,61}
\definecolor{justred}{RGB}{201,24,40}
\definecolor{deepred}{RGB}{111,3,25}
\usepackage[colorlinks,
linkcolor=red,
anchorcolor=blue,
citecolor=green
]{hyperref}

\DeclareMathOperator{\Tr}{Tr}
\DeclareMathOperator{\Rank}{Rank}

\begin{document}

\preprint{APS/123-QED}

\title{When can a local Hamiltonian be recovered from a steady state?}

\author{Jing Zhou}

\affiliation{Institute of Physics, Beijing National Laboratory for Condensed
  Matter Physics,\\Chinese Academy of Sciences, Beijing 100190, China}
	
\affiliation{School of Physical Sciences, University of Chinese Academy of
  Sciences, Beijing 100049, China}

\author{D. L. Zhou} \email[]{zhoudl72@iphy.ac.cn}
	
\affiliation{Institute of Physics, Beijing National Laboratory for Condensed
  Matter Physics,\\Chinese Academy of Sciences, Beijing 100190, China}
	
\affiliation{School of Physical Sciences, University of Chinese Academy of
  Sciences, Beijing 100049, China} \affiliation{Collaborative Innovation Center
  of Quantum Matter, Beijing 100190, China}
	
\affiliation{Songshan Lake Materials Laboratory, Dongguan, Guangdong 523808,
  China}

\date{\today}

\begin{abstract}

  With the development of quantum many-body simulator, Hamiltonian tomography
  has become an increasingly important technique for verification of quantum
  devices. Here we investigate recovering the Hamiltonians of two spin chains
  with $2$-local interactions and $3$-local interactions by measuring local
  observables. For these two models, we show that when the chain length reaches a
  certain critical number, we can recover the local Hamiltonian from its one
  steady state by solving the homogeneous operator equation (HOE) developed in
  Ref.~\cite{PhysRevLett.122.020504}. To explain the existence of such a
  critical chain length, we develop an alternative method to recover Hamiltonian
  by solving the energy eigenvalue equations (EEE). By using the EEE method, we
  completely recovered the numerical results from the HOE method. Then we
  theoretically prove the equivalence between the HOE method and the EEE method.
  In particular, we obtain the analytical expression of the rank of the
  constraint matrix in the HOE method by using the EEE method, which can be used
  to determine the correct critical chain length in all the cases.

\end{abstract}

\maketitle


\section{INTRODUCTION}\label{sec:introduction}

In quantum mechanics, all the information on a quantum system is contained in
its Hamiltonian~\cite{Projected, Resonating}. For example, all the energy
eigenvalues and eigenstates can be obtained by solving the eigenvalue problem of
the Hamiltonian. For a condensed matter system, however, its (effective)
Hamiltonian is unknown when prepared. Then it is crucial to determine the
Hamiltonian by making some quantum measurements, which is called Hamiltonian
tomography~\cite{PhysRevLett.102.187203,Lundeen2009,PhysRevA.80.022333}. In a
Hamiltonian tomography, the quantum measurements made must provide sufficient
information such that the Hamiltonian can be specified uniquely. For a generic
Hamiltonian, a successful Hamiltonian tomography needs the information on all
the energy eigenvalues and eigenstates. This implies that the number of
independent quantum measurements increases exponentially with the number of
particles in a Hamiltonian tomography.

Fortunately, the Hamiltonian of a physically realizable system is usually not
generic but local, which means that interactions arise only between (or among)
local particles~\cite{PhysRevLett.91.210401, PhysRevB.91.045138}. This
information on the local interaction pattern of the Hamiltonian is extremely
useful to reduce the necessary information from the quantum measurements in a
Hamiltonian tomography. For example, it has been shown that the local
Hamiltonian can be reconstructed uniquely by the information on one eigenstate
when the particle number becomes large in most cases~\cite{Does,Qi}.

One of the major challenges in Hamiltonian tomography is to develop an algorithm
to recover the Hamiltonian from numbers of measurements which is in accord with
demands of resource limitation with high accuracy~\cite{2021Sample}. The
measurement resource in the state-of-the-art algorithm for recovering a generic
local Hamiltonian scales polynomial to the system size~\cite{PhysRevX.9.041011,
Optimized}. Many algorithms have been proposed to recover the Hamiltonian by
making quantum measurements on its eigenstate~\cite{PhysRevLett.122.020504,
Hou_2020, Cao_2020, PhysRevB.99.235109, PhysRevLett.122.150606,
PhysRevA.86.022339}, dynamics~\cite{PhysRevLett.113.080401,8022944,
PhysRevLett.112.190501, PhysRevA.79.020305} and quantum
quench process~\cite{PhysRevLett.124.160502}. Several algorithms have been employed to
successfully recover some local Hamiltonians with a specific
pattern~\cite{Bairey_2020, PhysRevB.100.134201,Kokail2021}.

Recent years have witnessed the rapid development of quantum simulators and
computation devices, such as controlling trapped ions~\cite{2011Quantum,
  PhysRevLett.121.180501, PhysRevLett.117.060504, PhysRevLett.74.4091} and
superconducting circuits~\cite{PhysRevLett.111.080502,Barends2014}. To verify
the above devices, it's necessary to recover its Hamiltonian from the measured
observables, which makes the Hamiltonian tomography become increasingly
important in condensed matter physics and quantum computing. Given the practical
value of Hamiltonian tomography and significant development of numerical methods
of this task, several Hamiltonian tomography algorithms haven been implemented
on real physical systems~\cite{HOU2017863, PhysRevA.103.042429, Experimental}.

However, there is still a fundamental problem in Hamiltonian tomography that
does not have a satisfactory answer: When can a local Hamiltonian be uniquely
recovered from a steady state? Note that this problem have been solved
partially. For example, the authors in Ref.~\cite{PhysRevLett.122.020504} found
that when the rank of the constraint matrix equals to the number of independent
parameters minus one, the Hamiltonian can be uniquely recovered. However, we do
not know what factors determine the rank of the constraint matrix for a given
local Hamiltonian. Here we aim to present an analytical answer to this
fundamental problem. Based on our analytical results, in particular, we can
predict the critical chain length for any local Hamiltonian, i.e., the
Hamiltonian with the chain length beyond which can be uniquely recovered.

The paper is organized as follows. In Sec.~\ref{sec:reconstr-hamilt-homo}, after
reviewing the HOE method derived from a steady
state~\cite{PhysRevLett.122.020504}, we implement the HOE to reconstruct two
local Hamiltonians in a spin chain from a steady state, where we find that HOE
fails to recover Hamiltonian when the chain length is smaller than the critical
chain length. In Sec.~\ref{sec:determ-rank-constr}, we develop the EEE method to
give the analytical expressions for the critical chain lengths for any local
Hamiltonian tomography. Sec.~\ref{sec:determ-rank-constr} contains three
subsections. In Sec.~\ref{sec:reconstr-hamilt-ener}, we employ the EEE method to
recover the same two local Hamiltonians as studied in
Sec.~\ref{sec:reconstr-hamilt-homo}. In Sec.~\ref{sec:equiv-betw-hoe}, we prove
the equivalence of the HOE and EEE in a Hamiltonian tomography. In
Sec.~\ref{sec:determ-rank-constr-1}, we determine when the local Hamiltonian can
be recovered from a steady state. In Sec.~\ref{sec:conclusion}, we give a brief
summary.

\section{Reconstructing Hamiltonian by homogeneous operator equations}
\label{sec:reconstr-hamilt-homo}

In this section, we review and apply the method developed in
Ref.~\cite{PhysRevLett.122.020504} to solve the Hamiltonian tomography problem
whose goal is to recover the Hamiltonian of a quantum system by measuring some
observables when the system stays in a steady state. According to quantum
mechanics, the steady state may be an eigenstate of the Hamiltonian, or a mixed
state of several such eigenstates.

In general, the Hamiltonian to be recovered is decomposed as the sum
\begin{equation}
  \label{eq:1}
  H = \sum_{n=1}^{N} a_{n} h_{n}, 
\end{equation} 
where each $h_{n}$ is a known Hermitian operator, $a_{n}$ is an unknown real
parameter and $N$ is the number of terms in the Hamiltonian. The task of the
Hamiltonian tomography is to determine the vector
$\vec{a}=(a_{1},a_{2},\cdots,a_{N})$, which is formed by all the unknown parameters
in the Hamiltonian. Suppose that our system stays in the steady state $\rho$. Thus
the expectation value of any observable $K$ is invariant under the quantum
dynamic evolution, which is expressed in the Heisenberg picture as
\begin{equation}
  \label{eq:2}
  \partial_{t}\langle K\rangle = -\langle i[K,H]\rangle = 0, 
\end{equation}
where $\langle O\rangle=\Tr[O\rho]$ denotes the expectation value of the observable
$O$ in the steady state $\rho$. Inserting Eq.~\eqref{eq:1} into Eq.~\eqref{eq:2},
we conclude a homogeneous linear equation for the vector $\vec{a}$,
\begin{equation}
  \label{eq:3}
  \sum_{n=1}^{N} a_{n} \langle i[K,h_{n}]\rangle = 0. 
\end{equation}
Since Eq.~\eqref{eq:3} works for any observable $K$, we can choose a set of
observables $\{K_{m}\}_{m=1}^{M}$ and obtain $M$ linear constraints on the
vector $\vec{a}$
\begin{equation}
  \label{eq:4}
  \forall m: \sum_{n=1}^{N}a_{n}\langle i[K_{m},h_{n}]\rangle=0, 
\end{equation} 
which can be briefly written in the matrix form as
\begin{equation}
  \label{eq:5}
  G\vec{a}=0,\quad G_{mn}=\langle i[K_{m},h_{n}]\rangle. 
\end{equation}
Eqs.~\eqref{eq:4} or Eqs.~\eqref{eq:5} are called the linear homogeneous
operator equations (HOE), which are the basic equations to recover the
Hamiltonian developed in Ref.~\cite{PhysRevLett.122.020504}.

The degree of freedom of the vector $\vec{a}$ satisfying Eqs.~\eqref{eq:5} is
determined by the rank of the constraint matrix $G$, denoted as $\Rank G=r$. In
the Hamiltonian tomography, we assume that there always exists a nonzero
solution of $\vec{a}_{\text{true}}$, which implies that the rank $r< N$. The
rank of $G$ larger, the solutions of $\vec{a}$ more determined. However, even
when the rank of $G$ arrives at its maximum $r=N-1$, there still are an infinite
number of solutions in the form of $\alpha\vec{a}_{\text{true}}$ with $\alpha$ being any
real number. To remove the trivial ambiguity of the solutions, we reconstruct
the task into a convex optimization with constraint
\begin{equation}
  \label{eq:6}
  \min_{\vec{a}}||G\vec{a}||,  \text{ s.t. } ||\vec{a}||=1. 
\end{equation} 

The solution to Eq.~\eqref{eq:6} is the lowest right-singular vector of the
constraint matrix $G$, i.e., the row vector of $V^{T}$ that corresponds to the
lowest singular value of $G$ in the singular value decomposition
$G = U\Sigma V^{T}$. The error of the reconstructing task is defined as the distance
between the normalized true vector $\vec{a}_{\text{true}}$ and the recovered
vector $\vec{a}_{\text{recovered}}$
\begin{equation}
  \label{eq:7}
  \Delta =
  \left\Arrowvert\frac{\vec{a}_{\text{true}}}{||\vec{a}_{\text{true}}||} - \frac{\vec{a}_{\text{recovered}}}{||\vec{a}_{\text{recovered}}||}\right\Arrowvert. 
\end{equation} 

In Ref.~\cite{PhysRevLett.122.020504}, the HOE method has been applied to
recover the local Hamiltonian from local measurements. More precisely, the local
Hamiltonians of 6 middle qubits in a one-dimensional $12$-qubit chain with
random two-local interactions are successfully recovered by measuring the middle
qubits.

Here, we apply the HOE to study how to recover the local Hamiltonian from one
single steady state. For comparison, we study recovering the Hamiltonians of two
forms of spin $1/2$ chain. The first spin chain consists of local terms and
nearest-neighbor interactions, whose Hamiltonian
\begin{equation}
  \label{eq:8}
  H_{2} = \sum_{l=1}^{L} \sum_{\eta} a_{l\eta} \sigma_{l}^{\eta} + \sum_{l=1}^{L-1}
  \sum_{\eta} \sum_{\theta} a_{l\eta\theta} \sigma_{l}^{\eta} \sigma_{l+1}^{\theta},
\end{equation}
where $L$ is the spin chain length, $\eta$ and $\theta$ take values in the set
$\{x,y,z\}$, $\sigma_{l}^{\eta}$ is the $\eta$ component of the Pauli matrix of the
$l$-th spin, and all $a_{l\eta}$ and $a_{l\eta\theta}$ are the unknown parameters to be
recovered.

The second spin chain consists all three-neighbor interactions besides the terms
appearing in the Hamiltonian $H_{2}$, i.e., its Hamiltonian
\begin{equation}
  \label{eq:9}
  H_{3} = H_{2} + \sum_{l=1}^{L-2} \sum_{\eta} \sum_{\theta} \sum_{\delta} a_{l\eta\theta\delta} \sigma_{l}^{\eta} \sigma_{l+1}^{\theta}
  \sigma_{l+2}^{\delta}. 
\end{equation}
The state prepared to be measured is the mixed state which is a mixture of $q$
eigenstates of $H$, the Hamiltonian to be recovered.

The corresponding single steady state is assumed to be
\begin{equation}
  \label{eq:10}
  \rho = \sum_{j=1}^{q} p_{\mu}|\lambda_{\mu}\rangle\langle\lambda_{\mu}|,
\end{equation}
where $|\lambda_{\mu}\rangle$ is the $\mu$-th eigen state of the Hamiltonian
($H_{2}$ or $H_{3}$) with nonzero probability $p_{\mu}$, and $q$ is the rank of
the state $\rho$. Intuitively, the Hamiltonian $H_{3}$ contains more unknown
parameters than the Hamiltonian $H_{2}$, and we expect that $H_{3}$ is more
difficult to be recovered from the information in a steady state.

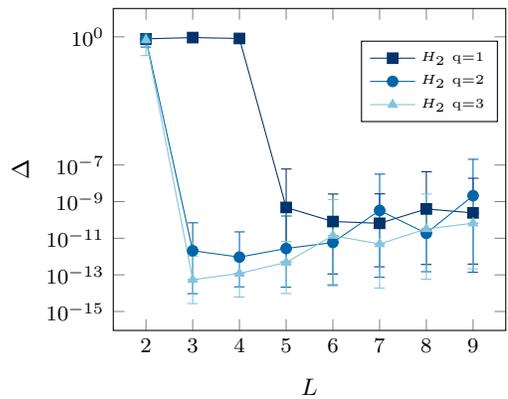
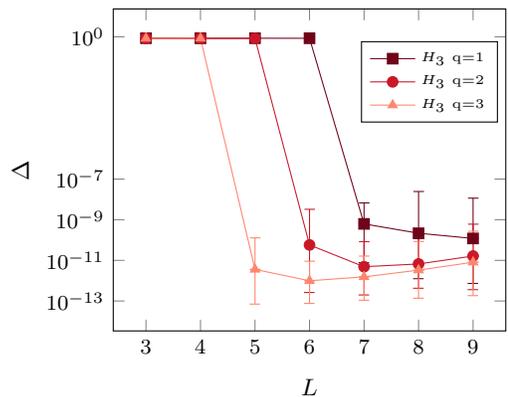
\begin{figure}[h]

  \subfigbottomskip=10pt \subfigure[]{
    \begin{tikzpicture}[align=center]
      \begin{semilogyaxis}
	[ small, width=6.8cm, xlabel = $L$, ylabel=$\Delta$, ytick style = {right},
        ytick={0,1e-17,1e-15,1e-13,1e-11,1e-9,1e-7,1},
        xtick={1,2,3,4,5,6,7,8,9}, legend style = {at={(0.98,0.9)}}, legend
        entries = {[font=\tiny]$H_{2}$ q=1,[font=\tiny]$H_{2}$
          q=2,[font=\tiny]$H_{2}$ q=3}]
	
	\addplot[mark=square*,deepblue, error bars/.cd, x dir=both, x explicit,
        y dir=both, y explicit, ] coordinates {
          (2,0.7812404383742485)+=(0,0.4008226844713282)-=(0,0.40297110841527123)
          (3,0.9120851056361005)+=(0,0.22736680038114232)-=(0,0.328955352733145)
          (4,0.8171437492213759)+=(0,0.27496126565539336)-=(0,0.29823312752004205)
          (5,4.820723635701528e-10)+=(0,6.002815982233946e-08)-=(0,4.815666977273865e-10)
          (6,8.179808751458903e-11)+=(0,2.516413854105564e-09)-=(0,8.168613484701032e-11)
          (7,6.55193584139026e-11)+=(0,2.603780684154165e-09)-=(0,6.524446668906643e-11)
          (8,3.9774204852555335e-10)+=(0,4.261083314540848e-08)-=(0,3.9737494371988615e-10)
          (9,2.4222138518738524e-10)+=(0,1.8707646293869747e-08)-=(0,2.418332822443608e-10)
        }; \addplot[mark=*,justblue, error bars/.cd, x dir=both, x explicit, y
        dir=both, y explicit, ] coordinates { (2,0.7784911489026896) +=
          (0,0.3999514702521354) -= (0,0.5022922426819413)
          (3,2.084544658824667e-12)+=(0,6.76741484399332e-11)
          -=(0,2.0751355481147906e-12)
          (4,9.32352212782835e-13)+=(0,2.1306516495649422e-11)
          -=(0,9.101814548916118e-13)
          (5,2.7494252389599483e-12)+=(0,1.6089693839805033e-10)
          -=(0,2.7281899508234073e-12)
          (6,5.888739786213039e-12)+=(0,1.2587091998041775e-10)
          -=(0,5.862236970130725e-12)
          (7,3.297766849835811e-10)+=(0,3.1569315848049024e-08)
          -=(0,3.2970264650727197e-10)
          (8,1.858509597197405e-11)+=(0,3.93201099458889e-10)
          -=(0,1.84363069135056e-11) (9,2.1156429989256418e-09)
          +=(0,2.0335006396863254e-07)-=(0,2.1155041358356773e-09) };
        \addplot[mark=triangle*,babyblue, error bars/.cd, x dir=both, x
        explicit, y dir=both, y explicit, ] coordinates { (2,0.6793695955535926)
          += (0,0.5353385131611514) -= (0,0.5841578351410246)
          (3,5.44576272212014e-14)+=(0,1.0063446694041803e-12)
          -=(0,5.177144838836887e-14)
          (4,1.1978142204622561e-13)+=(0,6.859752337623904e-13)
          -=(0,1.136471699557149e-13)
          (5,4.763759755647054e-13)+=(0,6.281210071058867e-12)
          -=(0,4.668513086703614e-13)
          (6,1.397038448166893e-11)+=(0,1.2852174801614904e-09)
          -=(0,1.3940597704541033e-11)
          (7,4.870767144109112e-12)+=(0,2.1128739421475437e-10)
          -=(0,4.851717619293476e-12)
          (8,3.237421286362211e-11)+=(0,2.548121936297528e-09)
          -=(0,3.2317004209957045e-11) (9,6.621218257448893e-11)
          +=(0,1.863989007894375e-09) -=(0,6.599845649326736e-11) };
      \end{semilogyaxis}
    \end{tikzpicture}
  }

  \subfigure[]{
    \begin{tikzpicture}[align=center]
      \begin{semilogyaxis}
	[ small, width=6.8cm, xlabel = $L$, ylabel=$\Delta$, ytick style = {right},
        ytick={0,1e-17,1e-15,1e-13,1e-11,1e-9,1e-7,1},
        xtick={1,2,3,4,5,6,7,8,9}, legend style = {at={(0.98,0.9)}}, legend
        entries = {[font=\tiny]$H_{3}$ q=1,[font=\tiny]$H_{3}$
          q=2,[font=\tiny]$H_{3}$ q=3}] \addplot[mark=square*,deepred, error
        bars/.cd, x dir=both, x explicit, y dir=both, y explicit, ] coordinates
        {
          (3,0.8531803911722051)+=(0,0.26951504984169317)-=(0,0.1824569090541448)
          (4,0.8440876295300771)+=(0,0.15339969555319533)-=(0,0.12788268870908792)
          (5,0.8508028159419989)+=(0,0.11967030315041038)-=(0,0.09590773448475998)
          (6,0.8475346412718817)+=(0,0.11342936022757244)-=(0,0.11544014048967921)
          (7,6.219383766642163e-10)+=(0,6.08750547003353e-09)-=(0,6.142305214410162e-10)
          (8,2.1566294719424427e-10)+=(0,2.4247887021124803e-08)-=(0,2.14411379502604e-10)
          (9,1.1938386467176926e-10)+=(0,1.1379436714399478e-08)-=(0,1.1865199009442253e-10)
	
	}; \addplot[mark=*,justred, error bars/.cd, x dir=both, x explicit, y
        dir=both, y explicit, ] coordinates { (3,0.841510086825776) +=
          (0,0.19741636427472942) -=(0,0.15156850010840572)
          (4,0.8357095740005803) += (0,0.14879640684275164)
          -=(0,0.13360149432010415) (5,0.8350887064149584) +=
          (0,0.13497060662692284) -=(0,0.09146253648967317)
          (6,5.729363745969538e-11) += (0,3.1965202899562106e-09)
          -=(0,5.703163037957464e-11) (7,4.852634467020476e-12) +=
          (0,7.872333286577332e-11) -=(0,4.654431556379819e-12)
          (8,6.641938557158194e-12) += (0,2.8398801431437953e-10)
          -=(0,6.218312695320738e-12) (9,1.6303447820856498e-11) +=
          (0,5.855376253862586e-10) -=(0,1.5943115767998955e-11)
	
	}; \addplot[mark=triangle*,babyred, error bars/.cd, x dir=both, x
        explicit, y dir=both, y explicit, ] coordinates { (3,0.831355425050025)
          += (0,0.1523542532040285) -=(0,0.17219067632020402)
          (4,0.8378581925442488) += (0,0.12062715634875465)
          -=(0,0.18243699226487742) (5,3.58212056975915e-12) +=
          (0,1.2379591788353567e-10) -=(0,3.511771381944668e-12)
          (6,9.903396730151498e-13) += (0,8.11582061932795e-12)
          -=(0,9.144931768966406e-13) (7,1.5384638896995085e-12) +=
          (0,1.4687938318072415e-11) -=(0,1.4292279568672318e-12)
          (8,3.3002329312220078e-12) += (0,8.181337050648859e-11)
          -=(0,3.1632661891362226e-12) (9,8.07387319444562e-12) +=
          (0,2.622541085854426e-10) -= (0,7.890594825689125e-12) };
      \end{semilogyaxis}
    \end{tikzpicture}
  }\\
  \caption{ We reconstruct (a) $H_{2}$ and (b) $H_{3}$ by means of HOE from
    steady state. Simulations are executed over $200$ random Hamiltonians with
    three different states for each chain length $L$. The squares, circles and
    triangles represent states of q=1,2 and 3, respectively. }\label{fig:1}
\end{figure}

\begin{table}[!t]
  \begin{minipage}[!t]{\columnwidth}
    \renewcommand{\arraystretch}{1.3} \centering \setlength{\tabcolsep}{0.6mm}{
      \begin{tabular}{|c|c|c|c|c|c|c|c|}
        \hline
        \quad & \quad & \multicolumn{2}{c|}{q=1} &  \multicolumn{2}{c|}{q=2}&  \multicolumn{2}{c|}{q=3}\\
        \hline
        \  L\ \ &\ N\ \  &\ \ r\ \ \  & \ \ $ \delta$\ \ &\ \  r\ \ \ & \ \ $\delta$ \ \ &\ \ r\ \ \ & \ \ $\delta$ \ \  \\
        \hline
        2&15 &6   & 8              &10 & 4              &12 & 2\\
        \hline
        3&27 &14 & 12           &26 & 0              &26 & 0\\
        \hline
        4&39 &30 & 8             &38 & 0              &38  & 0\\
        \hline
        5&51 &50 & 0             &51 & 0              & 51 & 0\\
        \hline
        6&63 &62 & 0             &62 & 0              &62  & 0\\
        \hline
        7&75 &74 & 0             &74 & 0              &74  & 0\\
        \hline
        8&87 &86 & 0             &86 & 0              &86  & 0\\
        \hline
        9&99 &98 & 0             &98 & 0              &98  & 0\\
        \hline
      \end{tabular}} 
    \caption{Reconstructing $H_{2}$ by means of HOE from steady state. $N$, $r$
      and $(N-1)-r$ with $q=1,2,3$ as the function of $L$.}\label{table:1}
  \end{minipage}
  \\[12pt]

\begin{minipage}[!t]{\columnwidth}
  \renewcommand{\arraystretch}{1.3}

  \centering \setlength{\tabcolsep}{0.8mm}{
    \begin{tabular}{|c|c|c|c|c|c|c|c|}
      \hline
      \quad & \quad & \multicolumn{2}{c|}{q=1} &  \multicolumn{2}{c|}{q=2}&  \multicolumn{2}{c|}{q=3}\\
      \hline
      \  L\ \ &\ N\ \  &\ \ r\ \ \  & \ \ $ \delta$\ \ &\ \  r\ \ \ & \ \ $\delta$ \ \ &\ \ r\ \ \ & \ \ $\delta$ \ \  \\
      \hline
      3&63 &14    & 48           &26 & 36              &36 & 26\\
      \hline
      4&111 &30  & 80            &58 & 52              &84  & 26\\
      \hline
      5&159 &62   & 96           &122 & 36              & 158 & 0\\
      \hline
      6&207 &126 & 80           &206 & 0              &206  & 0\\
      \hline
      7&255 &254 & 0             &254 & 0              &254  & 0\\
      \hline
      8&303 &302 & 0             &302 & 0              &302  & 0\\
      \hline
      9&351 &350 & 0             &350 & 0              &350  & 0\\
      \hline
    \end{tabular}}
  \caption{Reconstructing $H_{3}$ by means of HOE from steady state. $N$, $r$
    and $\delta$ with $q=1,2,3$ as the function of $L$}\label{table:2}
\end{minipage}
\vspace{-0.1cm}
\end{table}

Now we apply the HOE method reviewed in Sec. II to recover the above
Hamiltonians, whose procedure is given as follows. First, for each given chain
length $L$ we prepare the Hamiltonians to be recovered by generating $200$
random vectors $\{\vec{a}_{\text{true}}\}$ of the Gaussian distribution with
zero mean and unit standard deviation. Second, for each random vector
$\vec{a}_{\text{true}}$ we numerically calculate the eigenstates of the prepared
Hamiltonian and construct three mixed states $\rho$ with $q=1,2,3$ as given in
Eq.~\eqref{eq:10}. Third, we choose the terms $\{h_{i}\}_{i=1}^{N}$ in the
prepared Hamiltonian as the observables $\{K_{m}\}$, and calculate the
constraint matrix $G$. Note that such a choice makes the number of equations
equal to the number of unknown parameters in the prepared Hamiltonian. Fourth,
we calculate the recovered vector $\vec{a}_{\text{recovered}}$ by the singular
decomposition of $G$, along with the error as given in Eq.~\eqref{eq:7}.

Following the above HOE procedure, we numerically obtain the reconstructing
errors of the Hamiltonians $H_{2}$ and $H_{3}$ from mixed states $\rho$ with the
chain length $L$ from $1$ to $9$ shown in Fig.~\ref{fig:1}. We find that the HOE method
successfully recovers the Hamiltonian $H_{2}$ when $L\ge L_{c}$ where (1)
$L_{c}=5$ when $q=1$ (2) $L_{c}=3$ when $q=2$ (3) $L_{c}=3$ when $q=3$, and it
successfully recovers the Hamiltonian $H_{3}$ when $L\ge L_{c}$ (1) $L_{c}=7$ when
$q=1$ (2) $L_{c}=6$ when $q=2$ (3) $L_{c}=5$ when $q=3$. Here the reconstructing
error $\Delta \simeq 1$ implies the failure of the HOE method, and
$\Delta< 10^{-6}$ implies the success of the HOE method.

Note that for a given type of Hamiltonians, the more eigenstates are contained
in mixed state $\rho$, i.e., $q$ is larger, the more Hamiltonians can be uniquely
recovered. In addition, it is easier to recover $H_{2}$ than to recover $H_{3}$
from a steady state with the same $q$.

As discussed in Sec.~\ref{sec:reconstr-hamilt-homo}, the condition for the
Hamiltonian to be successfully recovered by the HOE method is
$\delta=N-(r+1)=0$, where $N$ is the number of unknown parameters in the
Hamiltonian, and $r$ is the rank of the constraint matrix $G$. Here we
numerically verify that $\delta=0$ only when $L\ge L_{c}$ for the Hamiltonian
$H_{2}$ and $H_{3}$, which are shown in Table.~\ref{table:1} and
Table.~\ref{table:2} respectively. In fact, the number of unknown parameters in
the Hamiltonians can be directly counted. For the Hamiltonian $H_{2}$,
$N=12L-9$; for the Hamiltonian $H_{3}$, $N=39L-63$. From the numerical results,
we observe that $r$ depends not only on the Hamiltonian (including the length
$L$) but also on the rank $q$ of the steady state. However, we have no idea of
how to directly determine the analytical relation between $r$ and the varibles
$L$ and $q$ for a given Hamiltonian from the HOE method.

\section{Determining rank of constraint matrix with energy eigenvalue equations}
\label{sec:determ-rank-constr}

To determine the value of $\Rank G$, it is instructive to study the energy
eigenvalue equations (EEE). In this section, we first apply EEE to recover
Hamiltonians $H_{2}$ and $H_{3}$ from the mixed state with different rank. Then,
we prove the equivalence of HOE and EEE. Finally, we determine the value of
$\Rank G$ using the characteristics of EEE.

\subsection{Reconstructing Hamiltonians by Energy Eigenvalue Equation}
\label{sec:reconstr-hamilt-ener}

When our system stays in the steady state $\rho$ in Eq.~\eqref{eq:10}, the most
complete information about the state $\rho$ can be obtained through quantum
tomography. In general, we assume the spectrum of $\rho$ is not degenerate. Then
we can explicitly obtain every eigenstate $|\lambda_{\mu}\rangle$ and its
probability $p_{\mu}$. Since generally the probability $p_{\mu}$ contains no
information of the Hamiltonian, all the information of the Hamiltonian is
contained in the eigenstates $\{|\lambda_{\mu}\rangle\}$. Based on this
consideration, we develop the following approach to recover the Hamiltonian
directly based on the energy eigenvalue equation, which is briefly called the
EEE approach.

The energy eigenvalue equation of local Hamiltonian $H=\sum_{n=1}^{N}a_{n}h_{n}$
can be written as
\begin{equation}
  \label{eq:11}
  \sum_{n=1}^{N} a_{n} h_{n} |\lambda_{\mu}\rangle = \lambda_{\mu} |\lambda_{\mu}\rangle,
\end{equation}
where $|\lambda_{\mu}\rangle$ is the eigenstate with eigenvalue $\lambda_{\mu}$ appearing in
Eq.~\eqref{eq:10}. In a specific basis $\{|i\rangle\}$ Eq.~\eqref{eq:11} becomes
\begin{equation}
  \label{eq:12}
  \sum_{n=1}^{N}a_{n}\langle i|h_{n}|\lambda_{\mu}\rangle=\lambda_{\mu}\langle i|\lambda_{\mu}\rangle.
\end{equation}
Splitting Eq.~\eqref{eq:12} into the real and the imaginary part
\begin{subequations}
  \label{eq:17}
  \begin{align}
    \sum_{n=1}^{N}a_{n}\Re\langle i|h_{n}|\lambda_\mu\rangle - \lambda_\mu \Re\langle i|\lambda_\mu\rangle &=0,\\
    \sum_{n=1}^{N}a_{n}\Im\langle i|h_{n}|\lambda_\mu\rangle - \lambda_\mu \Im\langle i|\lambda_\mu\rangle &=0,\label{Zb} 
  \end{align} 
\end{subequations}
where $\Re z$ and $\Im z$ denotes the real and the imaginary part of complex number
$z$ respectively. Denoting $L$ as the chain length, we can get $q\cdot2^{L+1}$
homogeneous linear equations with the unknowns
$\vec{x}=(a_{1},\cdots,a_{N},\lambda_{1},\cdots,\lambda_{q})$, which can be written in the matrix form
as
\begin{equation}
  Q \vec{x}=0,\label{eq:14}
\end{equation}
where the constraint matrix $Q$ is a $q\cdot 2^{L+1}\times (N+q)$ matrix:
\begin{equation}
  Q=\left(
    \begin{matrix}
      \label{eq:15}
      \Re A_{1} & \Re B_{1}\\
      \Im A_{1} & \Im B_{1}\\
      \Re A_{2} & \Re B_{2}\\
      \Im A_{2} & \Im B_{2}\\
      \vdots & \vdots \\
      \Re A_{q} & \Re B_{q}\\
      \Im A_{q} & \Im B_{q}
    \end{matrix}
  \right),
\end{equation}
with $A_{\mu}$ being a $2^{L}\times N$ matrix:
\begin{equation}
  \label{eq:16}
  A_{\mu} =
  \begin{pmatrix}
    \langle 1|h_{1}|\lambda_{\mu}\rangle & \cdots & \langle 1|h_{N}|\lambda_{\mu}\rangle \\
    \vdots & \vdots & \vdots \\
    \langle 2^{L}|h_{1}|\lambda_{\mu}\rangle & \cdots & \langle 2^{L}|h_{N}|\lambda_{\mu}\rangle
  \end{pmatrix}
\end{equation}
and $B_{\mu}$ being a $2^{L}\times q$ matrix with nonzero elements in its
$\mu$-th column:
\begin{equation}
  \label{eq:28}
  B_{\mu} =
  \begin{pmatrix}
    0 & \cdots & 0 & - \langle 1|\lambda_{\mu}\rangle & 0 & \cdots & 0 \\
    \vdots & \vdots & \vdots & \vdots & \vdots & \vdots & \vdots \\
    0 & \cdots & 0 & - \langle 2^{L}|\lambda_{\mu}\rangle & 0 & \cdots & 0
  \end{pmatrix}.
\end{equation}
Here Eq.~\eqref{eq:14} plays the same role as Eq.~\eqref{eq:5} in the HOE
method. Similarly, we can solve the parameter vector $\vec{a}$ by the following
constraint optimization problem
\begin{equation}
  \label{eq:18}
  \min_{\vec{a}}||Q \vec{x}||,  \text{ s.t. } ||\vec{a}||=1. 
\end{equation}
The degree of freedom of the vector $\vec{x}$ is determined by rank of
constraint matrix $Q$, which is denoted as $\Rank Q = r^{\prime}$. The number of
unknowns of linear equations Eq.~\eqref{eq:14} is represented as $N^{'}=N+q$.

The procedure to apply the EEE to recover the local Hamiltonians is given as
follows. First, for each given chain length $L$ we prepare the Hamiltonians to be
recovered by generating $200$ random vectors $\{\vec{a}_{\text{true}}\}$ of the
Gaussian distribution with zero mean and unit standard deviation. Second, for
each random vector $\vec{a}_{\text{true}}$ we numerically calculate the
eigenstates of the prepared Hamiltonian and construct three mixed states $\rho$
with $q=1,2,3$ as given in Eq.~\eqref{eq:10}. Third, we extract the eigenstates
by eigendecomposition of density matrix $\rho$. Then we construct constraint matrix
$Q$ by Eq.~\eqref{eq:15} and calculate the $\Rank Q$. Fourth, we solve the
Eq.~\eqref{eq:18} using the least-squares method by NumPy function
numpy.linalg.lstsq and calculate the reconstructing errors.

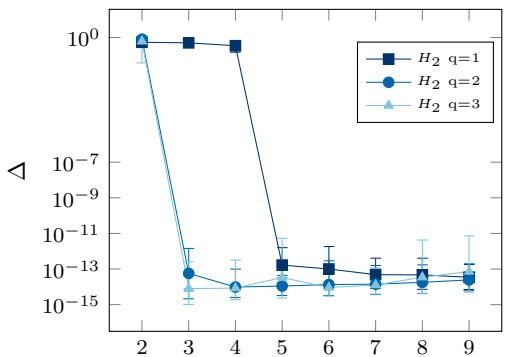
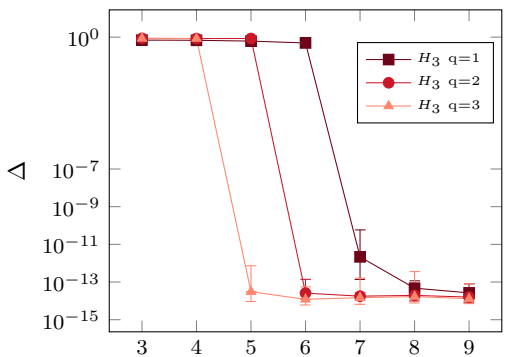
\begin{figure}[h]

  \subfigbottomskip=10pt \subfigure[]{
    \begin{tikzpicture}[align=center]
      \begin{semilogyaxis}
	[ small, width=6.8cm, xlabel = $L$, ylabel=$\Delta$, ytick style = {right},
        ytick={0,1e-17,1e-15,1e-13,1e-11,1e-9,1e-7,1},
        xtick={1,2,3,4,5,6,7,8,9}, legend style = {at={(0.98,0.9)}}, legend
        entries = {[font=\tiny]$H_{2}$ q=1,[font=\tiny]$H_{2}$
          q=2,[font=\tiny]$H_{2}$ q=3}]
	
	\addplot[mark=square*,deepblue, error bars/.cd, x dir=both, x explicit,
        y dir=both, y explicit, ] coordinates {

          (2,0.5399674772994085)+=(0,0.35982595334366385)-=(0,0.2402621482433121)
          (3,0.500865095587166)+=(0,0.24010982139246662)-=(0,0.20482681365388988)
          (4,0.3459996365363831)+=(0,0.18004155571426428)-=(0,0.19623456823192362)
          (5,1.6648059232421157e-13)+=(0,1.418662476394377e-12)-=(0,1.4989230840507287e-13)
          (6,1.0045467032817482e-13)+=(0,1.7338584390713384e-12)-=(0,9.196526507090234e-14)
          (7,4.851221593352513e-14)+=(0,3.5880734748050327e-13)-=(0,4.045529839854921e-14)
          (8,4.6608964431103454e-14)+=(0,3.585403180457076e-13)-=(0,3.932615020599642e-14)
          (9,3.441181943761035e-14)+=(0,1.4553784283897342e-13)-=(0,2.7346898390221976e-14)
        }; \addplot[mark=*,justblue, error bars/.cd, x dir=both, x explicit, y
        dir=both, y explicit, ] coordinates {

          (2,0.8041667467517922)+=(0,0.38113970406353925)-=(0,0.423341277699773)
          (3,5.702309711268152e-14)+=(0,1.3776257710917671e-12)-=(0,5.489084250363643e-14)
          (4,9.709305248384535e-15)+=(0,8.918272830298883e-14)-=(0,7.17309666306184e-15)
          (5,1.1093395566929887e-14)+=(0,3.285533632870053e-14)-=(0,7.844584525950721e-15)
          (6,1.3403638717706647e-14)+=(0,2.763341757664022e-13)-=(0,1.0148752928911672e-14)
          (7,1.4118304366961675e-14)+=(0,1.439698484622846e-13)-=(0,1.0416541916171387e-14)
          (8,1.7893773712397696e-14)+=(0,1.5418122968709286e-13)-=(0,1.3692150240422409e-14)
          (9,2.449297233752974e-14)+=(0,1.7216814926166886e-13)-=(0,1.8496426921013162e-14)
        }; \addplot[mark=triangle*,babyblue, error bars/.cd, x dir=both, x
        explicit, y dir=both, y explicit, ] coordinates {
	
          (2,0.613827308719193)+=(0,0.5394146409147373)-=(0,0.5763389676248031)
          (3,7.953932376862384e-15)+=(0,2.4682209399357187e-13)-=(0,6.931492569377754e-15)
          (4,8.500960221157288e-15)+=(0,3.069412060727657e-13)-=(0,6.581459103797606e-15)
          (5,3.342066520714807e-14)+=(0,5.287828785064855e-12)-=(0,3.1068331321168314e-14)
          (6,9.221905696155398e-15)+=(0,5.638916904914787e-14)-=(0,6.2132179234850194e-15)
          (7,1.258456780260412e-14)+=(0,1.0043737493222404e-13)-=(0,8.657039592005287e-15)
          (8,3.469730571492457e-14)+=(0,4.227357741064678e-12)-=(0,3.031133381435765e-14)
          (9,7.223807411608676e-14)+=(0,7.193549560682391e-12)-=(0,6.723172283535452e-14)
        };
      \end{semilogyaxis}
    \end{tikzpicture}
  }

  \subfigure[]{
    \begin{tikzpicture}[align=center]
      \begin{semilogyaxis}
	[ small, width=6.8cm, xlabel = $L$, ylabel=$\Delta$, ytick style =
        {right}, ytick={0,1e-17,1e-15,1e-13,1e-11,1e-9,1e-7,1},
        xtick={1,2,3,4,5,6,7,8,9}, legend style = {at={(0.98,0.9)}}, legend
        entries = {[font=\tiny]$H_{3}$ q=1,[font=\tiny]$H_{3}$
          q=2,[font=\tiny]$H_{3}$ q=3}] \addplot[mark=square*,deepred, error
        bars/.cd, x dir=both, x explicit, y dir=both, y explicit, ] coordinates
        {
	
          (3,0.6961942857835272)+=(0,0.13457755372768399)-=(0,0.16692134456857965)
          (4,0.680170703734136)+=(0,0.11900184775217293)-=(0,0.08501041674334742)
          (5,0.6158731075074886)+=(0,0.08940539525594626)-=(0,0.08890687632942562)
          (6,0.49063690423799644)+=(0,0.08396373973553428)-=(0,0.08115177629867376)
          (7,2.172884700562232e-12)+=(0,5.645786150130444e-11)-=(0,2.0349292230286606e-12)
          (8,4.728750863027432e-14)+=(0,6.890794265468611e-14)-=(0,2.6796059416838198e-14)
          (9,2.6622191160696524e-14)+=(0,5.326870119881579e-14)-=(0,1.7046284520817784e-14)
	
	}; \addplot[mark=*,justred, error bars/.cd, x dir=both, x explicit, y
        dir=both, y explicit, ] coordinates {
          (3,0.8629687427176056)+=(0,0.1935800090665063)-=(0,0.17512530998123)
          (4,0.8428322127449326)+=(0,0.16044160742136793)-=(0,0.11725948571887135)
          (5,0.847617434985672)+=(0,0.13074638963189944)-=(0,0.13062702871529108)
          (6,2.5780259261804118e-14)+=(0,1.1306337775285755e-13)-=(0,1.3877612688026088e-14)
          (7,1.7802500551368093e-14)+=(0,8.673636721397869e-15)-=(0,5.760452769375439e-15)
          (8,1.9555989053246063e-14)+=(0,4.8170269766930976e-14)-=(0,9.672744905170681e-15)
          (9,1.589623807075298e-14)+=(0,2.094211955891317e-14)-=(0,8.069544323604185e-15)
	
	}; \addplot[mark=triangle*,babyred, error bars/.cd, x dir=both, x
        explicit, y dir=both, y explicit, ] coordinates {
          (3,0.8493514012377125)+=(0,0.1819575998243469)-=(0,0.1659628283803336)
          (4,0.842123274723113)+=(0,0.1542887355936432)-=(0,0.12857392276289226)
          (5,3.0658760581522335e-14)+=(0,6.991001424761333e-13)-=(0,2.136359334191205e-14)
          (6,1.209171685621385e-14)+=(0,4.31896724238906e-14)-=(0,6.0372702832018455e-15)
          (7,1.479681105431732e-14)+=(0,1.4619722042660365e-13)-=(0,8.238232279055585e-15)
          (8,1.6914600607308742e-14)+=(0,3.4143371609655845e-13)-=(0,9.153487452776118e-15)
          (9,1.3044368974365172e-14)+=(0,6.909352596880223e-14)-=(0,5.8972318747468634e-15)
        };
      \end{semilogyaxis}
    \end{tikzpicture}
  }\\
  \caption{ We reconstruct (a) $H_{2}$ and (b) $H_{3}$ by means of EEE from
    steady state. Simulations are executed over $200$ random Hamiltonians with
    three different states for each chain length $L$. The squares, circles and
    triangles represent states of q=1,2 and 3, respectively. }\label{fig:2}
\end{figure}

\begin{table}[!t]
  \begin{minipage}[!t]{\columnwidth}
    \renewcommand{\arraystretch}{1.3} \centering \setlength{\tabcolsep}{0.6mm}{
      \begin{tabular}{|c|c|c|c|c|c|c|c|c|c|}
        \hline
        \quad  & \multicolumn{3}{c|}{q=1} &  \multicolumn{3}{c|}{q=2}&  \multicolumn{3}{c|}{q=3}\\
        \hline
        \ \ $L$\ \ & \ \ $N^{'}$\ \ &\ \ $r^{\prime}$ \ \  &\ \ $\delta^{'}$ \ \ &\ \ $N^{'}$ \ \ &\ \  $r^{\prime}$ \ \ & \ \ $\delta^{'}$ \ \ &\ \ $N^{\prime}$\ \  &\ \ $r^{\prime}$  \ \ & \ \ $\delta^{'}$ \ \ \\
        \hline
        2&16 &7   & 8         &17     &12     & 4      &18        &15 & 2\\
        \hline
        3&28 &15 & 12       &29    &28     & 0       &30       &29 & 0\\
        \hline
        4&40 &31 & 8        &41     &40     & 0       &42       &41  & 0\\
        \hline
        5&52 &51 & 0       &53      &52     & 0       &54       & 53 & 0\\
        \hline
        6&64 &63 & 0        &65     &64     & 0       &66       &65  & 0\\
        \hline
        7&76 &75 & 0        &77     &76     & 0       &78       &77  & 0\\
        \hline
        8&88 &87 & 0        &89     &88     & 0       &90       &89  & 0\\
        \hline
        9&100 &99 & 0     &101    &100    & 0       &102       &101  & 0\\
        \hline
      \end{tabular}} 
    \caption{Reconstructing $H_{2}$ by means of EEE from steady state. $N^{'}$,
      $r^{'}$ and $\delta^{'}$ with $q=1,2,3$ as the function of
      $L$.}\label{table:3}
  \end{minipage}
  \\[12pt]

\begin{minipage}[!t]{\columnwidth}
  \renewcommand{\arraystretch}{1.3}

  \centering \setlength{\tabcolsep}{0.8mm}{
    \begin{tabular}{|c|c|c|c|c|c|c|c|c|c|}
      \hline
      \quad  & \multicolumn{3}{c|}{q=1} &  \multicolumn{3}{c|}{q=2}&  \multicolumn{3}{c|}{q=3}\\
      \hline
      \ \ $L$\ \ & \ \ $N^{'}$\ \ &\ \ $r^{\prime}$ \ \  &\ \ $\delta^{'}$ \ \ &\ \ $N^{'}$ \ \ &\ \  $r^{\prime}$ \ \ & \ \ $\delta^{'}$ \ \ &\ \ $N^{\prime}$\ \  &\ \ $r^{\prime}$  \ \ & \ \ $\delta^{'}$ \ \ \\

      \hline
      3&64 &15 & 48       &65    &28     & 36       &66       &39  & 26\\
      \hline
      4&112 &31 & 80        &113     &60     & 52       &114       &87  & 26\\
      \hline
      5&160 &63 & 96       &161      &124     & 36       &162       & 161 & 0\\
      \hline
      6&208 &127 & 80        &209     &208     & 0       &210       &209  & 0\\
      \hline
      7&256 &255 & 0        &257     &256     & 0       &258       &257  & 0\\
      \hline
      8&304 &303 & 0        &305     &304     & 0       &306       &305  & 0\\
      \hline
      9&352 &351 & 0     &353    &352    & 0       &354       &353  & 0\\
      \hline
    \end{tabular}} 
  \caption{Reconstructing $H_{3}$ by means of EEE from steady state.
    $N^{\prime}$, $r^{\prime}$ and $\delta^{'}$ with $q=1,2,3$ as the functions
    of $L$ and $q$.}\label{table:4}
\end{minipage}
\vspace{-0.1cm}
\end{table}

Following the above EEE procedure, we obtain the reconstructing errors of the
Hamiltonians $H_{2}$ and $H_{3}$ from mixed states with chain length $L$ from $1$
to $9$ shown in Fig.~\ref{fig:2}. We present the accurate value of number of
unknowns $N^{'}$, $\Rank Q$ and difference $\delta^{'}=N^{'}-(r^{\prime}+1)$ as the
function of $L$ with $q=1,2,3$ for $H_{2}$ and $H_{3}$ in
Table.~\ref{table:3} and Table.~\ref{table:4}, respectively. For all the cases
with the same $N$ and $q$, we observe that
\begin{align}
  \label{eq:19}
  \delta^{\prime} & = \delta, \\
  r^{\prime} & = r + q.\label{eq:20}
\end{align}
Eq.~\eqref{eq:19} implies that the EEE method and the HOE method has the same
power to recover the Hamiltonians in all the cases, which are numerically
verified by the results shown in Fig.~\ref{fig:1} and Fig.~\ref{fig:2}.
Eq.~\eqref{eq:20} show that the rank of $G$ can be obtained by calculating the
rank of $Q$.

Up to now, we tackled the Hamiltonian tomography problem of $H_{2}$ and $H_{3}$
by both HOE and EEE methods. It turns out that the Local Hamiltonians space that
HOE and EEE can successfully recover contains the area that satisfies
$\delta=\delta^{\prime}=0$. In other words, it gives the same critical chain length
$L_{c}$ in all the cases. Here, we emphasize that, in the HOE procedure, when
all the Hamiltonian terms are used as observables $\{K_{m}\}$, adding new
observables to matrix $G$ will not increase the value of $\Rank G$.
Subsequently, $r$ in Table.~\ref{table:1} and Table.~\ref{table:2} is the
maximum value of $\Rank G$ to the corresponding type of Hamiltonians. We can
infer that the number of linearly independent functions in HOE can be no more
than the number of independent functions in EEE.


\subsection{Equivalence between HOE and EEE}
\label{sec:equiv-betw-hoe}

In this subsection, we prove the equivalence of HOE and EEE.

We first derive the HOE from the EEE.

In the EEE method, the complex conjugation of Eq.~\eqref{eq:12} gives
\begin{equation}
  \label{eq:30}
  \sum_{n=1}^{N}a_{n}\langle\lambda_{\mu}|h_{n}|j\rangle = \lambda_{\mu}\langle\lambda_{\mu}|j\rangle,
\end{equation}
where $|j\rangle$ is any basis vector. Combining Eq.~\eqref{eq:12} and
Eq.~\eqref{eq:30}, for any two basis vectors $|i\rangle$ and $|j\rangle$ we obtain
\begin{equation}
  \label{eq:31}
  \langle \lambda_{\mu}|[|j\rangle\langle i|, H]|\lambda_{\mu}\rangle = 0,
\end{equation}
which immediately leads to the basic equations of the HOE:
\begin{equation}
  \label{eq:32}
  \sum_{n=1}^{N}a_{n}\langle i[K_{m},h_{n}]\rangle = \sum_{\mu,j,i}  i p_{\mu} \langle j|K_{m}|i\rangle \langle
  \lambda_{\mu}|[|j\rangle\langle i|, H]|\lambda_{\mu}\rangle = 0,
\end{equation}
where $K_{m}$ is any linear operator on the Hilbert space.

Now we derive the EEE from the HOE.

We start from the basic equations of the HOE, Eq.~\eqref{eq:32}. Because $K_{m}$
is an arbitrary operator, we can always make the coefficients
$p_{\mu} \langle j|K_{m}|i\rangle $ linear independent when
$\{p_{\mu}, \mu=1,2,\cdots,q\}$ are non-degenerate. Thus we obtain Eq.~\eqref{eq:31} from
Eq.~\eqref{eq:32}. Note that Eq.~\eqref{eq:31} can be written as
\begin{equation}
  \label{eq:33}
  \Tr(|j\rangle \langle i| [H, |\lambda_{\mu}\rangle\langle \lambda_{\mu}|]) = 0.
\end{equation}
Because $\{|j\rangle\langle i|\}$ constructs a basis of the operator space,
Eq.~\eqref{eq:33} gives
\begin{equation}
  \label{eq:34}
  [H, |\lambda_{\mu}\rangle\langle \lambda_{\mu}|] = 0,
\end{equation}
which implies that $|\lambda_{\mu}\rangle$ is an eigenstate of $H$, i.e., that it satisfies
the eigenvalue equation~\eqref{eq:11}. This completes our proof of the
equivalence of the HOE and the EEE.

Consequently, Eq.~\eqref{eq:19} and Eq.~\eqref{eq:20} directly follow from the
above equivalence.

\subsection{Determining the Rank of Constraint Matrix}
\label{sec:determ-rank-constr-1}

We are now in a position to determine $\Rank G$ from $\Rank Q$ by
Eq.~\eqref{eq:20}. As mentioned above, the matrix $Q$ satisfies
Eq.~\eqref{eq:14}, which contains $q\cdot2^{L+1}$ homogeneous linear equations with
the unknowns $\vec{x}=(a_{1},\cdots,a_{N},\lambda_{1},\cdots,\lambda_{q})$. However, Eq.~\eqref{eq:12}
gives
\begin{equation}
  \label{eq:28}
  \sum_{n} a_{n} \langle\lambda_{\nu}|h_{n}|\lambda_{\mu}\rangle = \lambda_{\mu}\delta_{\mu\nu}, \quad\mu,\nu =1,\cdots,q
\end{equation}
Since $\langle\lambda_{\nu}|h_{n}|\lambda_{\mu}\rangle$ is complex in general, the above equations gives
$2q^{2}$ real constraint linear equations. Because every $h_{n}$ is Hermitian,
Eq.~\eqref{eq:28} implies
\begin{equation}
  \label{eq:35}
  \sum_{n} a_{n} \langle\lambda_{\mu}|h_{n}|\lambda_{\nu}\rangle = \lambda_{\mu}\delta_{\mu\nu}, \quad\mu,\nu =1,\cdots,q
\end{equation}
Then there are $q^{2}$ constraint independent homogeneous linear equations with
the unknowns $\vec{x}$ in Eq.~\eqref{eq:28}. Thus there are at most
$q\cdot2^{L+1}-q^{2}$ independent linear equations in Eq.~\eqref{eq:14}, i..e.,
$r^{\prime}\le q\cdot2^{L+1}-q^{2}$. In addition, because there are always nonzero
solutions of Eq.~\eqref{eq:14}, which implies that $r^{\prime}\le N+q-1$. Therefore, we
obtain
\begin{equation}
  \label{eq:36}
  r^{\prime}=\min\{q\cdot2^{L+1}-q^{2},N+q-1\}.
\end{equation}
By using Eq.~\eqref{eq:20}, we arrives at
\begin{equation}
  \label{eq:37}
  r=\min\{q\cdot2^{L+1}-q^{2}-q,N-1\}.
\end{equation}
The above analytical expressions of the rank of $G$ in Eq.~\eqref{eq:37} and the
rank of $Q$ in Eq.~\eqref{eq:36} are numerically verified in
Tables~\ref{table:1},\ref{table:2},\ref{table:3},\ref{table:4}.

The critical chain length is denoted as $L_{c}$. When chain length
$L\geq L_{c}$, we can uniquely recover the corresponding Hamiltonian. Now, we
determine the $L_{c}$ from Eq.~\eqref{eq:37}. To uniquely recover the
Hamiltonian, $\Rank G$ should equal to the number of unknowns minus 1, which
leads to
\begin{equation}
  \label{eq:38}
  q\cdot2^{L+1}-q^{2}-q\geq N-1.
\end{equation}

For the $2$-local Hamiltonian $H_{2}$, $N=12L-9$. From Eq.~\eqref{eq:38}, the
critical chain length
\begin{equation}
  \label{eq:13}
  L_{c}(H_{2},\rho) = \min_{L} q\cdot2^{L+1}-q^{2}-q \ge 12 L  -10,
\end{equation}
where $L\ge 2$ and $1\leq q \le2^{L}$.

Similarly, for the $3$-local Hamiltonian $H_{3}$, $N=39L-63$. From
Eq.~\eqref{eq:38}, the critical chain length
\begin{equation}
  \label{eq:21}
  L_{c}(H_{3},\rho) = \min_{L} q\cdot2^{L+1}-q^{2}-q \ge 39 L  -64,
\end{equation}
where $L\ge 3$ and $1\leq q \le2^{L}$.

We point out that our method not only works for local Hamiltonians $H_{2}$ and
$H_{3}$, it can also be used to predict $L_{c}$ for any one-dimensional spin
$1/2$ chain with local Hamiltonians. To demonstrate its effectiveness, we
calculate the $L_{c}$ of $H_{2}^{'}$, which contains the nearest and the next
nearest neighbor interaction,
\begin{equation}
  H_{2}^{'}=H_{2}+\sum_{l=1}^{L-2}
  \sum_{\eta} \sum_{\theta} a_{l\eta\theta} \sigma_{l}^{\eta} \sigma_{l+2}^{\theta}.
\end{equation}
For the Hamiltonian $H_{2}^{'}$, $N=21L-27$. The critical chain length can be
calculated by
\begin{equation}
  \label{eq:13}
  L_{c}(H_{2}^{'},\rho) = \min_{L} q\cdot2^{L+1}-q^{2}-q \ge 21 L  -28.
\end{equation}

The critical chain length for $H_{2}$, $H_{2}^{'}$ and $H_{3}$ with
$q=1,\cdots, 6$ are shown in Table~\ref{table:5}.

\begin{table}[!h]
  \begin{tabular}{|c|c|c|c|c|c|c|}
    \hline
    \diagbox{H}{$L_{c}$}{$q$} & 1 & 2  & 3 & 4 & 5&6 \\
    \hline
    $H_{2}$  &5 &3 & 3& 3 & 3&3\\
    \hline
    $H_{2}^{'}$  &6 &4 &3& 3 & 3&3 \\
    \hline
    $H_{3}$  &7 &6 &5& 4 & 4&3 \\
    \hline
  \end{tabular}
  \caption{The critical chain length $L_{c}$ for $H_{2}$, $H_{2}^{\prime}$ and $H_{3}$ with $q=1, \cdots, 6$.}\label{table:5}
\end{table}


\section{CONCLUSION}
\label{sec:conclusion}

We revisit the problem of reconstructing a local Hamiltonian when the system
stays in a steady state by measuring a collection of observables. Applying the
HOE method to the two spin chains with $2$-local interactions and $3$-local
interactions, we numerically find that only when the chain length $L$ is not
less than some critical chain length $L_{c}$ can we uniquely recover the
corresponding local Hamiltonian. The critical chain length $L_{c}$ depends not
only on the spin chain model, but also on the rank $q$ of the steady state.

To explain the underlying mechanism for the existence of the critical chain
length $L_{c}$, we observe that when the rank $r$ of the constraint matrix $G$
is not less than the number of unknown parameters in the recovered Hamiltonian
minus $1$, the Hamiltonian can be uniquely recovered. To further determine the
rank $r$, we develop an alternative method called the EEE method, which is used
to recover all the results from the HOE method. Further more, we proved the
equivalence between the HOE method and the EEE method. Especially, we obtain the
analytical expression of the rank $r$ by using the EEE method, which can be used
to determine the critical chain length $L_{c}$ analytically.

Our work studies the condition for a local Hamiltonian can be recovered from its
one steady state. For the two spin chain models with $2$-local interactions and
$3$-local interactions, we show the Hamiltonians can be reconstructed uniquely
only when the chain length is not less than the critical chain length.
Furthermore, our quantitative method Eq.~\eqref{eq:38} for determining the
critical chain length $L_{c}$ can be used on any one-dimensional spin 1/2 chain
with local Hamiltonians. In principle, we can extend our analytical result on
the critical length to the critical system size for two-dimensional and
three-dimensional local Hamiltonians. We hope that our work will shed novel
light on the Hamiltonian tomography problem.

\begin{acknowledgements}
  This work is supported by NSF of China (Grants No. 11775300 and No. 12075310),
  the Strategic Priority Research Program of Chinese Academy of Sciences (Grant
  No. XDB28000000), and the National Key Research and Development Program of
  China (Grant No. 2016YFA0300603).
\end{acknowledgements}

\bibliographystyle{apsrev4-2} \bibliography{Ham_tomo}

\end{document}